\documentclass[twocolumn,showpacs,preprintnumbers,prc,superscriptaddress]{revtex4-2}
\usepackage{graphicx,color}
\usepackage{float}
\usepackage{bm}
\usepackage[pdftex,colorlinks=true, linkcolor = blue, citecolor=blue,urlcolor=blue, bookmarksnumbered=true, bookmarksopen=true]{hyperref}
\begin{document}

\title{Solving one-dimensional penetration problem for fission channel in the statistical Hauser-Feshbach theory}

\author{T. Kawano}
\email{kawano@lanl.gov}
\affiliation{Los Alamos National Laboratory, Los Alamos, NM 87545, USA}

\author{P. Talou}
\affiliation{Los Alamos National Laboratory, Los Alamos, NM 87545, USA}

\author{S. Hilaire}
\affiliation{CEA, DAM, DIF, F-91297 Arpajon, France}
\affiliation{Universit\'{e} Paris-Saclay, CEA, LMCE, 91680 Bruy\`{e}res-le-Ch\^{a}tel, France}

\date{\today}
\preprint{LA-UR-23-25089}

\begin{abstract}
We solve the Schr\"{o}dinger equation for an arbitrary one-dimensional
potential energy to calculate the transmission coefficient in the
fission channel of compound nucleus reactions. We incorporate the
calculated transmission coefficients into the statistical
Hauser-Feshbach model calculation for neutron-induced reactions on
$^{235,238}$U and $^{239}$Pu. The one-dimensional model reproduces the
evaluated fission cross section data reasonably well considering the
limited number of model parameters involved. A resonance-like
structure appears in the transmission coefficient for a double-humped
fission barrier shape that includes an intermediate well, which is
understood to be a quantum mechanical effect in the fission channel.
The calculated fission cross sections for the neutron-induced
reactions on $^{235,238}$U and $^{239}$Pu all exhibit a similar
structure.
\end{abstract}
\maketitle

\section{Introduction}
\label{sec:introduction}
The statistical compound nucleus theory describes the probability for
a formed compound nucleus to decay into a channel $a$ by the partial
width $\Gamma_a$, and the Hauser-Feshbach theory~\cite{Hauser1952}
tells us that the energy-average of width $\langle \Gamma_a \rangle$
can be replaced by the optical model transmission coefficient $T_a$ in
the time-reverse process. This is intuitive for particle or
photon-induced reactions, as the interpretation reads the strength to
decay into the channel $a$ is proportional to the compound nucleus
formation probability from the same channel. For the fission channel,
however, the reverse process is not at all trivial. Several
approximations and models are then employed, which significantly
complicate the comparison and interpretation with experimental fission
cross-section data. Studies on the nuclear fission have a long
history, and comprehensive review articles of the fission calculation
are given by Bj\o{}rnholm and Lynn~\cite{Bjornholm1980},
Wagemans~\cite{Wagemans1991}, and more recently Talou and
Vogt~\cite{Talou2023}.

A traditional approach is to calculate a penetrability (transmission
coefficient) through the fission barrier by adopting the
semi-classical Wentzel–Kramers–Brillouin (WKB)
approximation~\cite{Hill1953}.  We often assume that one-dimensional
(1-D) potential energy forms a double-humped fission barrier shape,
which is predicted by the liquid drop model with the microscopic
(shell and pairing energies) corrections, and apply WKB to each of the
barriers separately. By decoupling these two fission barriers, an
effective (net) transmission coefficient $T_f$ through the whole
potential energy is calculated as
\begin{equation}
  T_f = \frac{T_A T_B}{T_A + T_B} \ ,
  \label{eq:effectiveTf}
\end{equation}
where $T_A$ and $T_B$ are the WKB penetrability through the barriers.
Obviously this treatment over-simplifies the fission penetration
problem, as it ignores potential wells between barriers, which gives
rise to the so-called class-II and class-III (in the triple humped
case) states. Some attempts were made in the past to calculate the
fission transmission coefficient by considering the potential well
between barriers. For example, Sin {\it et al.}~\cite{Sin2006,
Sin2016} defined a continuous fission barrier shape and applied WKB
for each segment to calculate the effective transmission
coefficient. Bouland, Lynn, and Talou~\cite{Bouland2013} implemented
the transition states in the class-II well, through which the
penetrability is expressed in terms of the $R$-matrix
formalism. Romain, Morillon, and Duarte~\cite{Romain2016} reported an
anti-resonant transmission due to the class-II and class-III
states. Some recent developments in the fission calculations are
summarized in Ref.~\cite{Talou2023}.

Segmentation of the potential energy along the nuclear elongation
axis, where the inner barrier, class-II state, outer barrier,
class-III states, $\ldots$, are aligned, still implies that the
penetration through the entire potential energy surface is obtained by
assembling its piece-wise components. Although limited to an
analytical expression of potential energy, Cramer and
Nix~\cite{Cramer1970} obtained an exact solution of wave function in
terms of the parabolic-cylinder functions for the double-humped
potential shape. Sharma and Leboeuf~\cite{Sharma1976} extended this
technique to the triple-humped potential barrier case. By solving the
Schr\"{o}dinger equation numerically, an extension of the Cramer-Nix
model to an arbitrary shape of 1-D potential energy is
straightforward. This was reported by Morillon, Duarte, and
Romain~\cite{Morillon2010} and by ourselves~\cite{LA-UR-15-24956},
where the effective transmission coefficient in
Eq.~(\ref{eq:effectiveTf}) is no longer involved.  The solution of
Schr\"{o}dinger equation for 1-D potential is, however, just one of
all the possible fission paths, whereas the dynamical fission process
takes place through any excited states on top of the fission barrier
in a strongly deformed compound nucleus. To calculate the actual
fission transmission coefficient that can be used in the
Hauser-Feshbach theory calculations, we have to take into account the
penetration through the excited states as well.

Eventually we describe the nuclear fission process from two extreme
point of views, namely the compound nucleus evolves through a fixed
albeit large number of fission paths, or the configuration is fully
mixed in the potential well so that the penetration through the
multiple barriers can be totally decoupled as in
Eq.~(\ref{eq:effectiveTf}).

Our approach follows the more general former case; the fission process
takes place along an eigenstate of the compound nucleus, which is
continuous along the nuclear deformation coordinate. In this paper, we
revisit the Cramer-Nix model and its extension to the arbitrary
potential energy shape, and introduce nuclear excitation to calculate
$T_f$. The obtained $T_f$ is used in the Hauser-Feshbach theory to
calculate the fission cross section, which can be compared with
available experimental data. We perform the cross-section calculations
for two distinct cases, the neutron-induced fission on $^{238}$U where
the total excitation energy is still under the fission barrier, and
that for $^{235}$U and $^{239}$Pu where the system energy is higher
than the barrier. In this paper we limit ourselves to the first-chance
fission only, where no neutron emission occurs prior to
fission. However, extension to the multi-chance fission process is not
complicated at all.

\section{Theory}
\label{sec:theory}
\subsection{Fission transmission coefficient for double-humped fission barrier}

\begin{figure}
 \resizebox{\columnwidth}{!}{\includegraphics{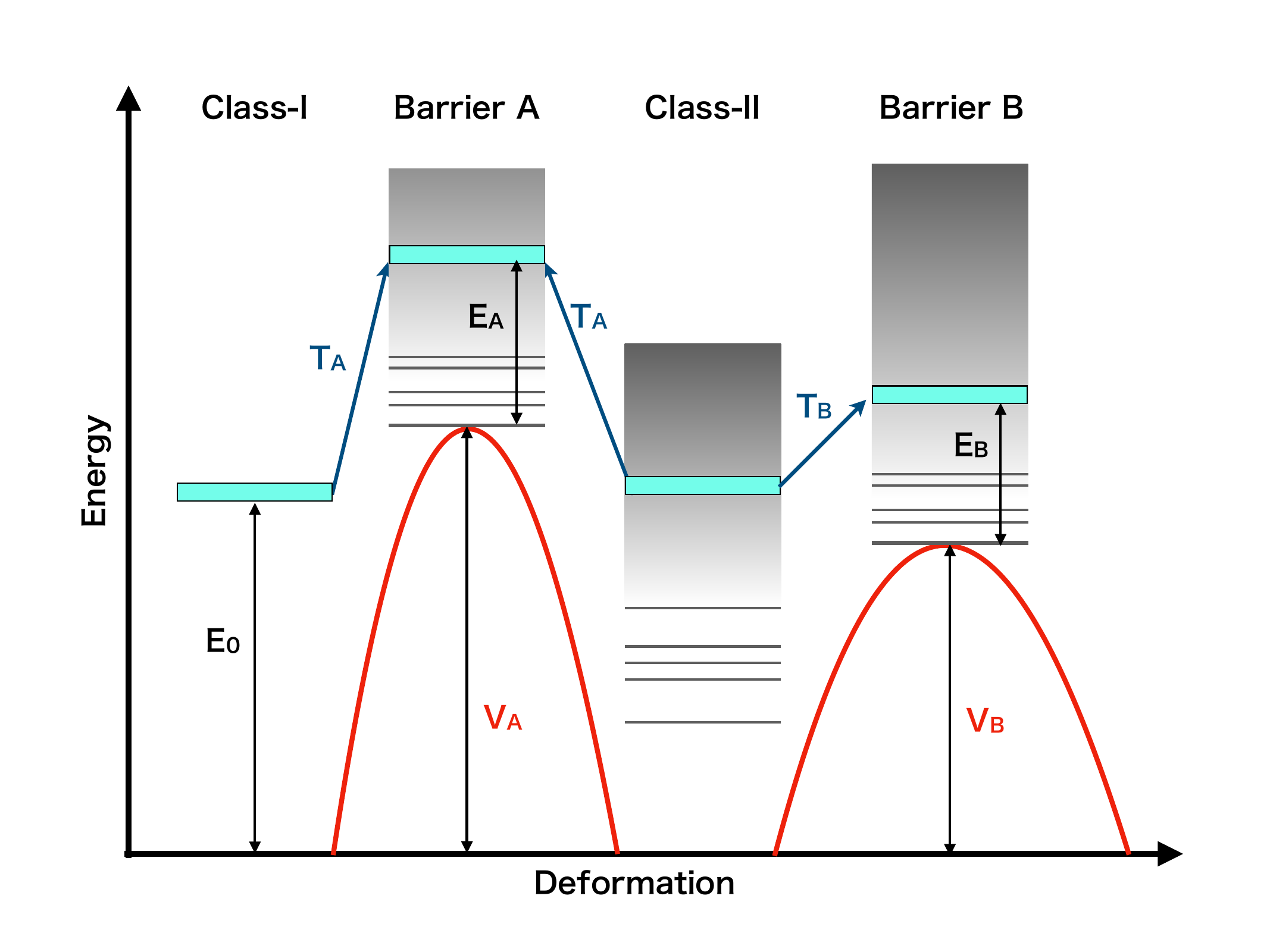}}
 \caption{Schematic picture of double-humped potential energy
   along the nuclear deformation direction, showing the double-humped
   fission barriers $V_A$ and $V_B$, and the class-I and class-II
   wells between the barriers. The initial compound nucleus state
   is at $E_0$ in class-I, which decays through the states at
   $E_A$ and $E_B$ on top of each barrier.}
 \label{fig:double_humped_barrier}
\end{figure}

First we briefly summarize the standard technique to calculate the
fission transmission coefficient $T_f$ for the double-humped fission
barrier. The objective is to emphasize the distinction between the
conventional fission calculation and our approach. The fission barrier
is approximated by an inverted parabola characterized by the barrier
parameters; the heights $V_A$ for the inner barrier and $V_B$ for the
outer barrier, and their curvatures $C_A$ and $C_B$ (the curvature is
often denoted by $\hbar\omega$), as shown schematically in
Fig.~\ref{fig:double_humped_barrier}. By applying the WKB
approximation to the parabolic-shaped barriers, the transmission
coefficient is given by the Hill-Wheeler expression~\cite{Hill1953}
\begin{equation}
  T_i(E)
  = {{1} \over
     {1 + \exp\left(
                2\pi \frac{V_i + E - E_0)}{C_i}
              \right)}} , \qquad i = A, B \ ,
  \label{eq:TfWKB}
\end{equation}
where $E_0$ is the initial excitation energy, $E$ is the nuclear
excitation energies measured from the top of each barrier.  The
``lumped'' transmission coefficient $T$ is the sum of all possible
excited states at $E_k$ for the discrete levels and at $E_x$ in the
continuum,
\begin{eqnarray}
  T_i
  &=& \sum_k T_i(E_k) \nonumber\\
  &+& \int_{E_c}^\infty T_i(E_x) \rho_i(E_x) dE_x , \qquad i = A, B \ ,
  \label{eq:Tlumped}
\end{eqnarray}
where $\rho(E_x)$ is the level density on top of each barrier, and
$E_c$ is the highest discrete state energy. Although we didn't specify
the spin and parity of the compound nucleus, the summation and
integration are performed for the same spin and parity states.  Often
some phenomenological models are applied to $\rho(E_x)$ to take the
nuclear deformation effect into account, which is the so-called
collective enhancement~\cite{Junghans1998}. A standard technique in
calculating fission cross sections, {\it e.g.} as adopted by
Iwamoto~\cite{Iwamoto2007}, assumes typical nuclear deformations at
the inner and outer barriers. Generally speaking the collective
enhancement is model and assumption dependent, which makes fission
model comparison difficult.

When the fission barriers $V_A$ and $V_B$ are fully decoupled, $T_A$
represents a probability to go through the inner barrier, and a
branching ratio from the intermediate state to the outer direction is
$T_B/(T_A+T_B)$.  The effective fission transmission coefficient is
thus given by Eq.~(\ref{eq:effectiveTf}). This expression implies that
the dynamical process in the class-II well is fully adiabatic, and it
virtually forms a semi-stable compound state.  It should be noted that
there is no explicit fission path in this model, since integration
over the excited states in Eq.~(\ref{eq:Tlumped}) is performed before
connecting $T_A$ and $T_B$.

\subsection{Fission transmission coefficient for 1-D shape}

\subsubsection{Concatenated parabolas}

\begin{figure}
 \resizebox{\columnwidth}{!}{\includegraphics{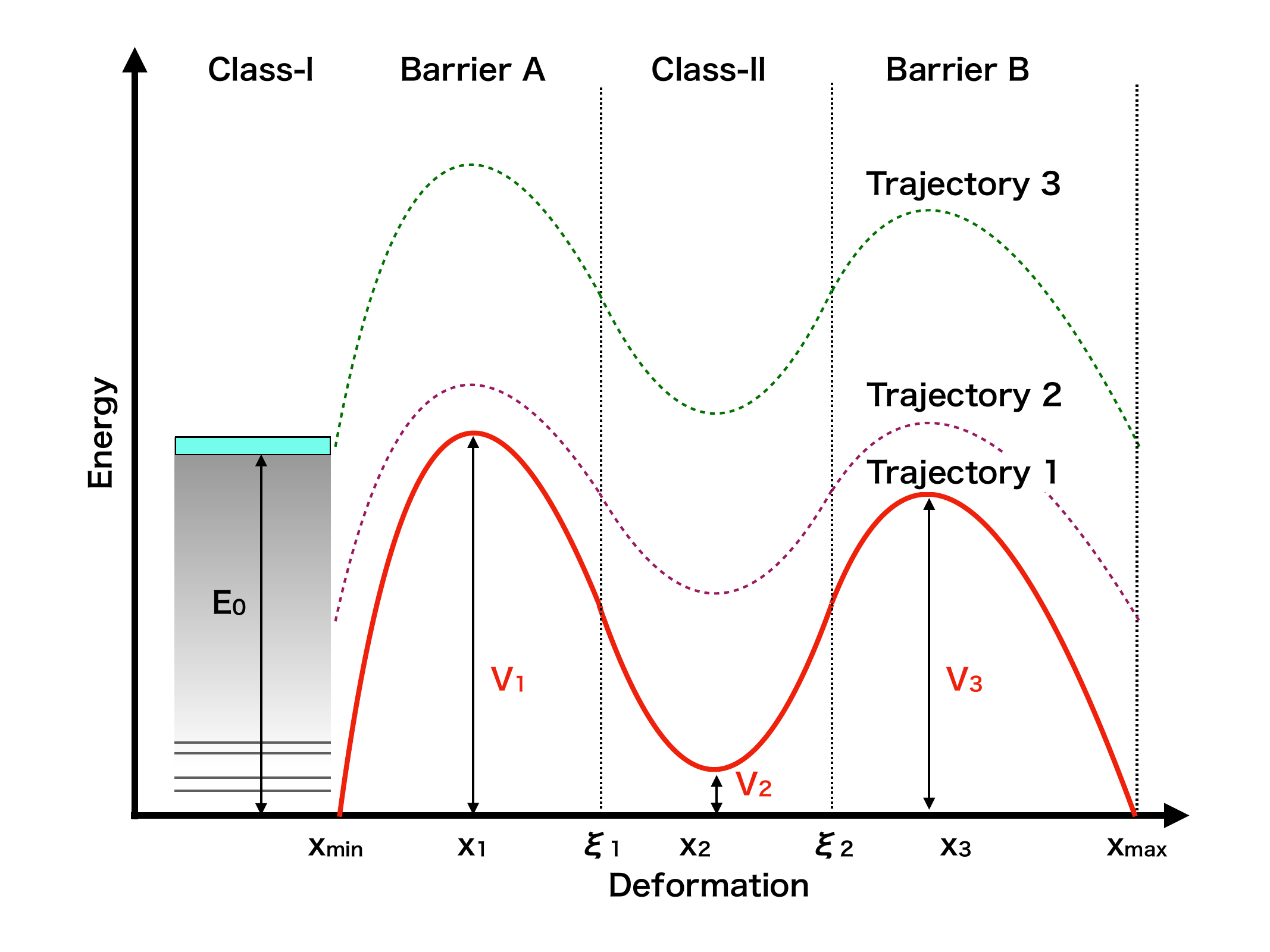}}
 \caption{Schematic picture of 1-D potential energy
   along the nuclear deformation direction. 
   The initial compound nucleus state is at $E_0$, 
   which decays through 1-D fission paths,
   {\it e.g.} the trajectories 1, 2, etc.}
 \label{fig:connected_barrier}
\end{figure}

The Schr\"{o}dinger equation for an arbitrary one-dimensional (1-D)
potential energy shape can be solved exactly without the WKB
approximation by applying the numerical integration
technique. Although our purpose is to solve problems for any fission
barrier shape, it is still convenient to employ the parabolic
representation to compare with the double-humped barrier
cases. Similar to the three-quadratic-surface parameterization of
nuclear shape~\cite{Nix1969, Moller2009}, the 1-D barrier is
parameterized by smoothly connected parabolas
\begin{equation}
  V(i,x) = V_i + (-1)^i \frac{1}{2} c_i (x - x_i)^2  , \qquad i = 1, 2, \ldots \ ,
  \label{eq:potential}
\end{equation}
where $i$ is the region index for the segmented parabola (odd $i$ for
barriers, and even for wells), $x$ is a dimensionless deformation
coordinate, $c_i = \mu C_i^2 / \hbar^2$, $V_i$ is the top (bottom)
energy of the barrier (well), $x_i$ is the center of each parabola,
and $\mu$ is the inertial mass parameter. Note that the region index
adopted here corresponds to the double-humped case as $A = 1$ and $B =
3$.  Because the deformation coordinate is dimensionless, the
calculated result is insensitive to $\mu$, and we take
\begin{equation}
  \frac{\mu}{\hbar^2} = 0.054 A^{5/3} \qquad \mbox{MeV$^{-1}$}
\end{equation}
as suggested by Cramer and Nix~\cite{Cramer1970}. The region index $i$
runs from 1 to 3 for the double-humped shape, and 5 for the
triple-humped shape. The double-humped case is shown in
Fig.~\ref{fig:connected_barrier} by the solid curve.

By providing the barrier parameters $V_i$ and $C_i$, the junction
point ($\xi_i$) and the parabola center ($x_i$) for each adjacent
region are automatically determined through continuity
relations. Since the abscissa is arbitrary in the 1-D model, we first
fix the center of the first barrier at
\begin{equation}
  x_1 = x_{\rm min} + \sqrt{ \frac{2 V_1}{c_1} } \ ,
\end{equation}
where $x_{\rm min}$ is an arbitrary small offset. The consecutive
central points are given by
\begin{equation}
  x_i = x_{i-1}
      + \sqrt{ \frac{2|V_{i-1} - V_i|(c_{i-1}+c_i)}{c_{i-1}c_i} } \ ,
  \label{eq:center}
\end{equation}
and the junction points are 
\begin{equation}
  \xi_i = \frac{c_i x_i + c_{i+1} x_{i+1}}{c_i + c_{i+1} } \ .
  \label{eq:junction}
\end{equation}
With the central points of Eq.~(\ref{eq:center}) and the junction
points of Eq.~(\ref{eq:junction}), the segmented parabolas in
Eq.~(\ref{eq:potential}) are smoothly concatenated.

In the class-II and/or class-III well between the barriers, it is
possible to add a small imaginary potential that accounts for flux
absorption~\cite{Sin2016}
\begin{equation}
  W(i,x) = 
  \left\{
  \begin{array}{ll}
    \Delta V - W_i & \Delta V \leq W_i \\
    0              & \Delta V  > W_i \\
  \end{array}
  \right. \ ,
\end{equation}
where $\Delta V = V(i,x) - V_i$, $i=2$ for class-II and $i=4$ for
class-III. We assume the potential shape is the same as the real part,
while the strength is given by a parameter $W_i$.

\subsubsection{Solution of 1-D Schr\"{o}dinger equation}

The 1-D Schr\"{o}dinger equation for the fission channel of compound
nucleus at the system energy $E$ is written
as~\cite{Cramer1970}
\begin{equation}
  \frac{d^2}{dx^2}\phi(x)
  + \frac{2\mu}{\hbar^2}
     \left\{ E - \left( V(x) + iW(x) \right) \right\} \phi(x) = 0 \ ,
\end{equation}
where the wave function $\phi(x)$ satisfies the following boundary
condition~\cite{Back1971}
\begin{equation}
  \phi(x) \simeq
  \left\{
  \begin{array}{ll}
    u^{(-)}(kx) - S u^{(+)}(kx) & x > x_{\rm max} \\
    A u^{(-)}(kx) & x < x_{\rm min} \\
  \end{array}
  \right. \ .
  \label{eq:boundary}
\end{equation}
$[x_{\rm min}, x_{\rm max}]$ is the entire range of fission
barrier considered, $k =\sqrt{2\mu E}$ is the wave number, $A$ is the
amplitude of wave function in the class-I well, and
\begin{equation}
  u^{(\pm)}(kx) = \cos(kx) \pm i \sin(kx) \ .
  \label{eq:extfunc}
\end{equation}
The Sch\"{o}dinger equation in the internal region can be solved
numerically by a standard technique such as the Numerov method or
Fox-Goodwin method~\cite{Fox1949}. The solution at the matching point
$x_m$ in the external region ($x_m > x_{\rm max}$) is written as
\begin{equation}
  \psi(x_m) = u^{(-)}(kx_m) - S u^{(+)}(kx_m) \ ,
\end{equation}
and the internal solution $\phi(x_m)$ is smoothly connected with the
external solution at $x_m$. Analog to the scattering matrix element in
the single-channel optical model, the coefficient $S$ is then given by
\begin{equation}
  S = \frac{f u^{(-)}(x_m) - g^{(-)}}{f u^{(+)}(x_m) - g^{(+)}} \ ,
\end{equation}
where 
\begin{equation}
  f \equiv \left. \frac{d\phi / dx}{\phi} \right|_{x_m} \quad \mbox{and}\quad
  g^{(\pm)} \equiv \left. \frac{d u^{(\pm)}}{dx} \right|_{x_m} \  .
\end{equation}

When the potential is real everywhere, the fission transmission
coefficient is given by
\begin{equation}
  T = 1 - |S|^2 \ .
\end{equation}
In the case of the complex potential, the amplitude $A$ in
Eq.~(\ref{eq:boundary}) is given by the normalization factor of the
internal wave function at $x_m$,
\begin{equation}
  A = \left. \frac{u^{(-)} - S u^{(+)}}{\phi} \right|_{x_m} \ ,
\end{equation}
and the transmission coefficient through the barrier is $T_d = |A|^2$.
Because of the loss of flux due to the imaginary potential, $T_d$ is
smaller than $T$, and $T_d$ goes into the statistical Hauser-Feshbach
theory instead of $T$.

\subsubsection{Potential energy for excited states}

Since penetration through the potential defined by
Eq.~(\ref{eq:potential}) is merely one of all the possible fission
paths, we have to aggregate such possible trajectories (paths) to
calculate the lumped transmission coefficient, which is analogous to
Eq.~(\ref{eq:Tlumped}). While the fission penetration for the ground
state takes place through the shape of potential energy in
Eq.~(\ref{eq:potential}), each of the excited states would be
constructed on top of the ground state trajectory. This is a critical
difference between the double-humped and 1-D models, as an adiabatic
intermediate state assumed in the double-humped model conceals an
actual fission path along the deformation coordinate, while it is
explicit in the 1-D model.

To define the fission trajectories for the excited states, one of the
most naive assumptions is that the potential energy is shifted by the
excitation energy $E_x$ as $V(x) = V_0(x) + E_x$, where $V_0$ is the
potential for the ground state.  This, however, ignores distortion of
the eigenstate spectrum in a compound nucleus as it changes shape.  At
the limit of adiabatic change in the nuclear shape, the excitation
energy of each of the eigenstates changes slightly due to shell,
pairing, and nuclear deformation effects.  This results in distortion
of the trajectories, as opposed to a simple shift in energy.

We empirically know that calculated fission cross sections
underestimate experimental data if we simply adopt the level density
$\rho(E_x)$ for an equilibrium shape in the lump-sum of
Eq.~(\ref{eq:Tlumped}). Therefore we often employ some models to
enhance the level densities on top of each of the barriers, which
account for increasing collective degree-of-freedom in a strongly
deformed nucleus. Instead of introducing the collective enhancement in
our 1-D penetration calculation, we assume the excitation energies of
the states will be lowered due to the nuclear deformation. In other
words, the eigenstates in a compound nucleus at relatively low
excitation energies are distorted by deformation effects.  An
illustration of the distortion effect corresponding to a compression
is schematically shown in Fig.~\ref{fig:connected_barrier} by the
dotted curves---trajectories 2 and 3. This trajectory compression
should be mitigated for the higher excitation energies, which is also
phenomenologically known as the damping of collectivity. Although the
compression might depend on the deformation as it changes the pairing
and shell effects, we model the compression in a rather simple way to
eliminate unphysical over-fitting to observed data. We assume the
eigenstates in the compound nucleus are compressed by a factor that
depends on the excitation energy only. Our ansatz reads
\begin{equation}
  \varepsilon_x =
      \left\{ 
        f_0 + \left(1 - e^{-f_1 E_x} \right) \left(1 - f_0\right)
      \right \} E_x , \
  \label{eq:compression}
\end{equation}
where the parameter $f_0$ is roughly 0.8 and the damping
$f_1$ is $\sim 0.2$~MeV$^{-1}$ as shown later. The corresponding fission
trajectory for the excited states is now
\begin{equation}
   V(x) = V_0(x) + \varepsilon_x \ .
\end{equation}
The transmission coefficient for this trajectory is
$T(\varepsilon_x)$, and the lumped transmission coefficient $T_f$ is
given by
\begin{equation}
  T_f  = \sum_k T(\varepsilon_k) 
       + \int_{E_c}^\infty T(\varepsilon_x) \rho(\varepsilon_x) dE_x \ ,
  \label{eq:Tlumped2}
\end{equation}
where the summation and integration are performed for the spin and
parity conserved states. Although the integration range goes to
infinity, or some upper-limit value could be
considered~\cite{Hilaire2003}, this converges quickly with increasing
excitation energy. Generally it is safe to truncate the integration at
$E_x = E_0$.

\section{Results and Discussion}
\label{sec:results}
\subsection{Wave function and transmission coefficient for a single fission path}

\begin{figure}
 \resizebox{\columnwidth}{!}{\includegraphics{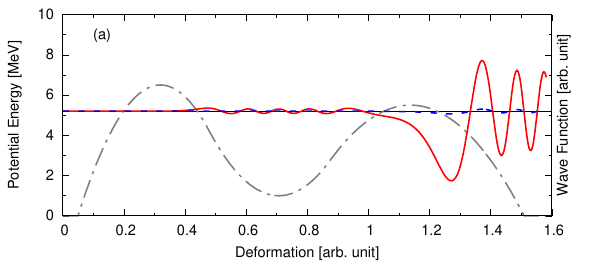}}\\
 \resizebox{\columnwidth}{!}{\includegraphics{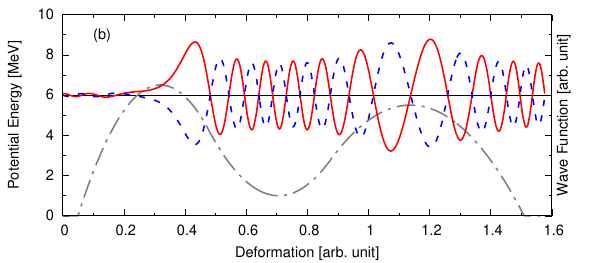}}\\
 \resizebox{\columnwidth}{!}{\includegraphics{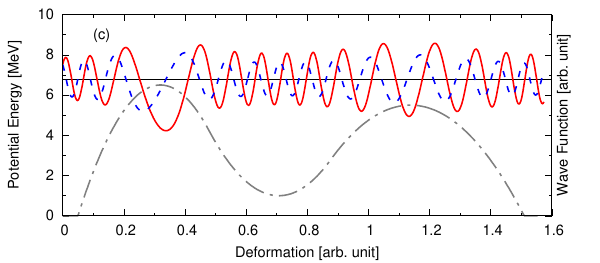}}
 \caption{Calculated wave functions for the connected parabolas.
   The potential energy is shown by the dot-dashed curves. The solid and dotted
   curves are the normalized wave function (solid for the real part,
   and dotted for the imaginary part). (a) the system energy lies
   below both barrier heights, (b) the energy is between the barriers,
   and (c) is above the barriers case.}
 \label{fig:wavefunc}
\end{figure}

As an example of the 1-D model, the calculated wave functions for
connected parabolas are shown in Fig.~\ref{fig:wavefunc}, which is for
the $A=239$ system. The assumed barrier heights are $V_1=6.5$,
$V_2=1$, and $V_3=5.5$~MeV, with the curvatures of $C_1=0.6$ and
$C_2=0.4$, and $C_3=0.5$~MeV. We depict the three cases of system
energy $E$; (a) below both of the barriers, (b) between $V_1$ and
$V_3$, and (c) above the both.

Since the 1-D potential penetration problem is invariant whether
numerical integration is performed from the right or left side, the
wave function is normalized to the external function that has unit
amplitude. The penetrability is seen as the amplitude of wave function
inside the potential region. Apparently the wave function penetrates
through the potential barrier when the system has enough energy to
overcome the both barriers $E>V_1$ and $E>V_3$, and it is blocked if
the barrier is higher than the system energy. However, although the
wave function damps rapidly, quantum tunneling is still seen beyond
the barrier.

One of the remarkable differences from the double-humped model in
Eq.~(\ref{eq:effectiveTf}) with the Hill-Wheeler expression of
Eq.~(\ref{eq:TfWKB}) is that the 1-D model sometimes exhibits
resonating behavior due to the penetration through the class-II
well. This was already reported by Cramer and Nix~\cite{Cramer1970} in
their parabolic-cylinder function expression. It should be noted that
this is not an actual compound nucleus resonance, but a sort of the
size effect where the traveling and reflecting waves have accidentally
the same phase. As a result the wave function is amplified
significantly at a resonating energy.

\begin{figure}
 \resizebox{\columnwidth}{!}{\includegraphics{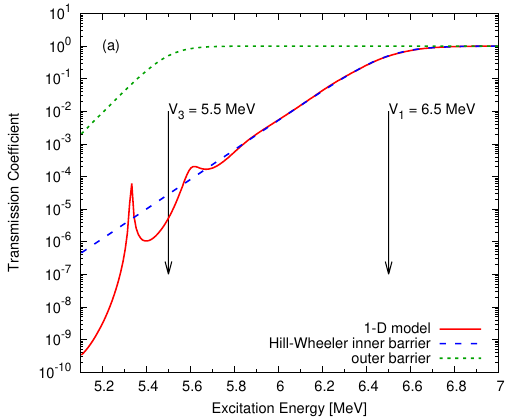}}\\
 \resizebox{\columnwidth}{!}{\includegraphics{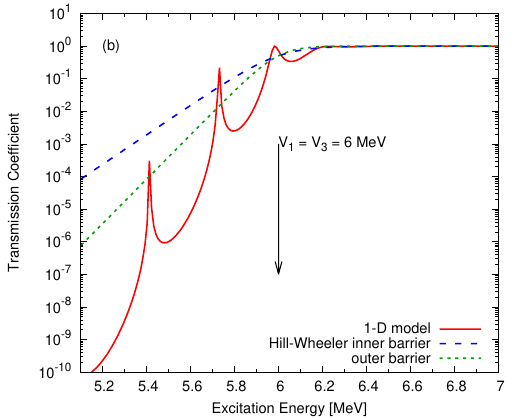}}
 \caption{Calculated transmission coefficients for the connected parabolas
   as a function of the excitation energy. The top panel 
   (a) is for the potential characterized by $V_1=6.5$, $V_2=1.0$,
   and $V_3=5.5$~MeV, with the curvatures of $C_1=0.6$ and $C_2=0.4$,
   and $C_3=0.5$~MeV, 
   and the bottom panel (b) is for the $V_1=V_3 = 6.0$~MeV case.
   The dashed and dotted curves are the WKB approximation for the inner and 
   outer barriers.}
 \label{fig:transmission}
\end{figure}

This amplification can be seen easily in the transmission coefficient
as in Fig.~\ref{fig:transmission}.  The top panel is for the same
potential as the one in Fig.~\ref{fig:wavefunc}.  The first sharp
resonance appears below the inner barrier of $V_1=5.5$~MeV, and the
second broader resonance is just above the barrier. We also depicted
the transmission coefficients calculated with the WKB approximation in
Eq.~(\ref{eq:TfWKB}) for the inner and outer barriers. As it is known,
the WKB approximation works reasonably well when the energy is close
to the fission barrier. However, it deviates notably from the 1-D
model when an interference effect of penetrations through the inner
and outer barriers becomes visible.

This effect becomes more remarkable when the inner and outer barriers
have a similar magnitude, which results in a special circumstance that
the penetration and reflection waves are in phase. The bottom panel in
Fig.~\ref{fig:transmission} is the case where these barriers have the
same height of 6.0~MeV. A broad resonance appears just below the
fission barrier, which enhances the fission cross section even if the
compound state is still below the fission barrier. Then the
penetration drops rapidly as the excitation energy decreases. On the
contrary, the penetration by WKB stays higher in the sub-threshold
region. Under these circumstances, the Hill-Wheeler expression may
give unreliable fission cross sections, albeit their magnitude would
be quite small. Nuclear reaction codes sometimes introduce a
phenomenological class-II (and class-III) resonance effect to
compensate for this deficiency~\cite{Talou2023, RIPL3}.

The difference in the WKB curves in Fig.~\ref{fig:transmission} (b) is
due to the curvatures, and both curves approach to $T_A = T_B = 1$
once the system has more than the barrier energy. However, the
effective transmission coefficient becomes 1/2, when
Eq.~(\ref{eq:effectiveTf}) is applied. This is also an important
difference between the double-humped and 1-D models, as the 1-D model
always gives $T = 1$ when the system energy can overcome all the
barriers.

\subsection{Fission path through complex potential}

\begin{figure}
 \resizebox{\columnwidth}{!}{\includegraphics{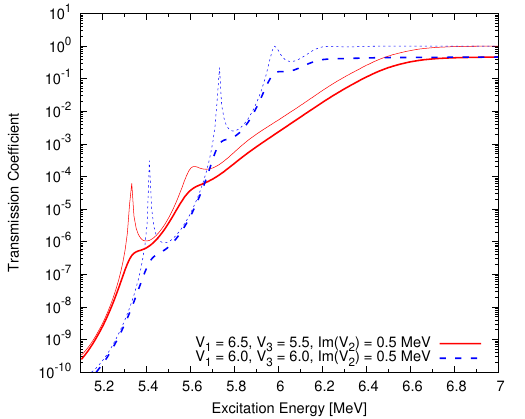}}
 \caption{Calculated transmission coefficients for complex potentials
   as a function of the excitation energy. The thin solid and dotted curves
   are the calculated transmission coefficients same as the one shown
   in Fig.~\ref{fig:transmission}. The thick curves are the transmission
   coefficients when an imaginary strength of 0.5~MeV is added to the
   class-II well.}
 \label{fig:transmissionImag}
\end{figure}

The wave function is absorbed by a potential when a complex class-II
well is given, which results in reduction in the fission transmission
coefficient. When we add a small imaginary part ($W_2 = 0.5$~MeV) to
class-II in the potentials shown in Fig.~\ref{fig:wavefunc}, the
calculated transmission coefficients are compared with the real
potential cases in Fig.~\ref{fig:transmissionImag}. The imaginary
potential acts on the wave function as amplitude damping so that the
asymptotic transmission coefficients at higher energies will be less
than unity. In this case, the asymptotic value of $T_d = |A|^2$ is
$\sim 0.5$, which is determined by $W_2$. The imaginary potential also
shifts the phase of wave function slightly, and the resonance-like
shape is less pronounced.

When a larger imaginary potential is provided, the fission
transmission coefficient goes to almost zero. A physical meaning of
the amplitude damping is not so definite, since the imaginary strength
is arbitrary. This is analogous to the optical model; an incident
particle disappears in the optical potential by its imaginary part
regardless of the nuclear reaction mechanisms. A possible
interpretation is that the system is trapped by a shape isomeric state
that might be long-lived.

\subsection{Hauser-Feshbach model calculation}

We incorporate the lumped fission transmission coefficient in
Eq.~(\ref{eq:Tlumped2}) into the statistical Hauser-Feshbach model
calculation to demonstrate applicability of the 1-D model in actual
compound nucleus calculations. We do not include the imaginary
potential, so that the calculated results will be tightly constrained
by the potential shape characterized by a limited number of inputs.

The calculation is performed with the CoH$_3$ statistical
Hauser-Feshbach code~\cite{Kawano2019}, which properly combines the
coupled-channels optical model and the statistical Hauser-Feshbach
theory by performing the Engelbrecht-Weidenm\"{u}ller
transformation~\cite{Engelbrecht1973, kawano2015, Kawano2016} of the
optical model penetration matrix~\cite{Satchler1963}. This is
particularly important for nuclear reaction modeling in the actinide
mass region. We employ the coupled-channels optical potential by
Soukhovitskii {\it et al.}~\cite{Soukhovitskii2004} for producing the
neutron penetration matrix and the generalized transmission
coefficients~\cite{Kawano2009}.

To look at the fission channel more carefully, we take some reasonable
model inputs for other reaction channels from literature and do not
attempt to make fine-tuning as the purpose of this study is not a
parameter fitting. Since the curvature parameter $C = \hbar\omega$ is
relatively insensitive to fission cross section calculation, we fix
them to a typical value of 0.6~MeV, and roughly estimate the heights
of inner and outer barriers as well as the trajectory compression
parameters in Eq.~(\ref{eq:compression}) by comparing with
experimental fission cross section data. The class-II depth has also a
moderate impact on the calculation of transmission coefficients as far
as we provide a reasonable value. We fix it to 0.5~MeV. Other model
parameters are set to default internal values in CoH$_3$.  The
$\gamma$-ray strength function is taken from Kopecky and
Uhl~\cite{Kopecky1990} with the M1 scissors mode~\cite{Mumpower2017},
the Gilbert-Cameron composite formula~\cite{Gilbert1965, Kawano2006}
for the level density, and the discrete level data taken from
RIPL-3~\cite{RIPL3}.

\begin{figure}
 \resizebox{\columnwidth}{!}{\includegraphics{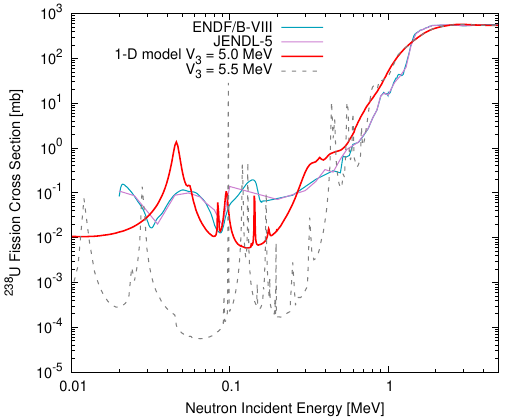}}
 \caption{Calculated fission cross section for neutron-induced reaction
   on $^{238}$U. The barrier height parameters are $V_1=6.0$, $V_2=0.5$,
   and $V_3=5.0$ for the solid curve. The dashed curve is for $V_3=5.5$~MeV.} 
 \label{fig:fissionU238}
\end{figure}

First, we perform the statistical model calculations for
neutron-induced reaction on $^{238}$U, where sub-threshold fission may
be seen below about 1~MeV of incident neutron energy.  The ground
state rotational band members, $0^+$, $2^+$, $4^+$, and $6^+$ are
coupled with the deformation parameters taken from the Finite Range
Droplet Model (FRDM)~\cite{Moller1995}. The calculated fission cross
sections are shown in Fig.~\ref{fig:fissionU238} by comparing with the
evaluated fission cross sections in ENDF/B-VIII.0~\cite{ENDF8} and
JENDL-5~\cite{JENDL5}.  The reason of showing the evaluations instead
of actual experimental data is that the evaluated data often include
more experimental information than the direct measurement of $^{238}$U
fission cross section, {\it e.g.}  cross section ratio
measurements. The accuracy of the evaluations is good enough to test
the relevance of this new model. We found that the case of $V_1=6$,
$V_2=0.5$, and $V_3=5$~MeV reasonably reproduce the evaluated fission
cross section in the energy range of our interest. The compression
parameters $f_0=0.8$ and $f_1=0.2$~MeV$^{-1}$ were needed to reproduce
the fission cross section plateau above 2~MeV.  We also calculated the
$V_3=5.5$~MeV case, which produces a more resonance-like structure below
1~MeV, despite the fact that they tend to underestimate the
evaluations on average.

Since the resonating behavior seen in the sub-threshold region
originates from the wave function in between the inner and outer
barriers, their locations and amplitudes strongly depend on the shape
of the potential energy surface.  Because the 1-D potential energy
constructed by smoothly concatenating segmented parabolas is a crude
approximation, we naturally understand that such structure in the
experimental data cannot be predicted exactly by the model unless we
modify the potential shape freely. This being said, the fission cross
sections calculated with the 1-D model in the sub-threshold region are
not so far from reality, which is usually not so obvious in the
Hill-Wheeler case.

\begin{figure}
 \resizebox{\columnwidth}{!}{\includegraphics{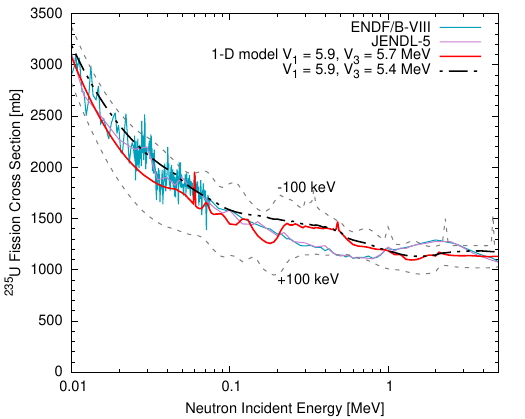}}
 \caption{Calculated fission cross section for neutron-induced reaction
   on $^{235}$U. The barrier height parameters are $V_1=5.9$, $V_2=0.5$,
   and $V_3=5.7$~MeV for the solid curve. The dashed curves are the case 
   when $V_1 \pm 100$~keV.} 
 \label{fig:fissionU235}
\end{figure}

The neutron-induced reaction on $^{235}$U does not have a threshold in
the fission channel. The neutron separation energy is 6.55~MeV, and
the compound nucleus already has enough energy to fission even for a
thermal-energy neutron incident. We adopt the same trajectory
compression parameters, $V_2$, and curvature parameters as those in
the $^{238}$U calculation, and just look for $V_1$ and $V_3$. We found
that the set of $V_1=5.9$ and $V_3=5.7$~MeV gives a reasonable fit to
the experimental $^{235}$U fission cross section, as compared with the
evaluated values in Fig.~\ref{fig:fissionU235}. The resonance-like
structure, which is seen in the sub-threshold fission of $^{238}$U, is
also seen near 60~keV. The evaluated data also show a small bump near
30~keV, which might be attributed to enhancement of the wave-function
amplitude in between the barriers. However, it is hard to claim that
our predicted peak at 60~keV corresponds to the observed bump, as the
potential energy shape is over-simplified in this study.

To show a sensitivity of the inner barrier (or the higher one), a
range of calculated fission cross sections by changing $V_1$ by $\pm
100$~keV is shown by the dashed curves.  More resonance-like structure
appears when $V_1$ is reduced to 5.8~MeV, because there is only a
100~keV difference between $V_1$ and $V_3$.  When the difference is
larger, $V_1+100$~keV, the structure becomes less pronounced.  A
similar sensitivity study was performed by Neudecker {\it el
  al.}~\cite{Neudecker2021}. where a 100--150~keV change in the fission
barrier height changes the calculated fission cross sections by 10\%
or so, while the cross section shape remains the same in the
conventional fission model.

While $V_1$ has such a large sensitivity, the outer barrier (or the
lower one) does not change the calculated fission cross section much,
as far as $V_3$ is lower than $V_1$ by a few hundred keV or
more. Figure~\ref{fig:fissionU235} includes the case of $V_1=5.9$ and
$V_3=5.4$~MeV, where the resonance-like structure is fully washed
out. We do not show the sensitivity of $V_3$ by further lowering the
outer barrier, since these curves are hard to distinguish
anymore. Astonishingly, the calculated fission cross sections remain
almost identical even if $V_3=1$~MeV, which implies that the fission
calculation is totally governed by the single-humped fission barrier
shape.

\begin{figure}
 \resizebox{\columnwidth}{!}{\includegraphics{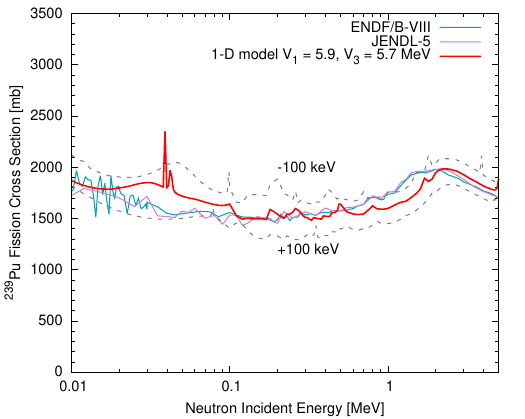}}
 \caption{Calculated fission cross section for neutron-induced reaction
   on $^{239}$Pu. The barrier height parameters are $V_1=5.9$, $V_2=0.5$,
   and $V_3=5.7$~MeV for the solid curve. The dashed curves are the case 
   when $V_1 \pm 100$~keV.} 
 \label{fig:fissionPu239}
\end{figure}

Figure~\ref{fig:fissionPu239} shows the calculated fission cross
section of $^{239}$Pu. In this case, it was difficult to obtain a
reasonable fit to the evaluations by employing the same compression
parameters, and a reduction of $f_0$ to 0.55 was needed ($f_1$ is the
same as before). The barrier height parameters are $V_1=5.9$,
$V_2=5.7$~MeV. The resonance-like structure also appears, although it
is not so noticeable like in the uranium cases. We also show the
cross-section band when $V_1\pm 100$~keV. The sensitivity of $V_1$ to
the fission cross section is similar to $^{235}$U. The evaluated cross
sections are roughly covered by the $\pm100$~keV band. However, again,
we emphasize that the objective of the present study is not to fit
perfectly the model calculation to the experimental data but to
demonstrate the fact that the simple 1-D model is potentially capable
of capturing the gross features of the fission reaction process by
producing calculated fission cross sections in reasonable agreement
with experimental data, without the need for a large number of fitting
model parameters.

\subsection{Possible refinements}

Although we employed the parameterized potential shape, which is
constructed by segmented parabolas, the experimental fission cross
sections are reasonably reproduced by a few model parameters that
characterize the shape itself. This is already a significant
improvement of the statistical Hauser-Feshbach calculations for
fission compared to the traditional Hill-Wheeler expression for the
double-humped fission barriers connected by
Eq.~(\ref{eq:effectiveTf}). For better reproduction of available
experimental data, as well as prediction of unknown fission cross
sections, we envision further improvement by incorporating a few
theoretical ingredients.

First, the potential energy shape could be taken from the potential
energy surface calculated microscopically~\cite{Goriely2009} or by
semi-microscopic approaches~\cite{Moller2009, Moller2015,
  Verriere2021, Jachimowicz2021}.  Because the potential energy
surface is often defined in a multi-dimensional deformation coordinate
space, either we have to project the surface onto a one-dimensional
axis (it is, however, known that the projection often causes
discontinuity problems~\cite{Dubray2012}), or our 1-D model should be
extended to a set of coupled-equations for the multi-dimensional
coordinate. Second, we should employ a better trajectory compression
model rather than the simple damping of Eq.~(\ref{eq:compression}),
where nuclear deformation effect is ignored, nevertheless it is known
that the single-particle spectrum depends on the nuclear
deformation. Because our trajectory compression model is constant
along the deformation axis, the potential penetration calculation
becomes invariant for exchange of the inner and outer barriers, while
the calculated potential energy surface often indicates that the inner
barrier tends to be higher than the outer barrier for the U and Pu
isotopes. Such a property might be seen by introducing the trajectory
distortion that is deformation dependent.  The nuclear
deformation can be calculated with the full- or semi-microscopic
approaches, where broken symmetries in the nuclear shape are naturally
taken into account. We could estimate possible trajectories by
calculating the microscopic level densities based on the single
particle energies in the deformed one-body potential.

\section{Conclusion}
\label{sec:conclusion}
We proposed a new model to calculate fission cross sections in the
statistical Hauser-Feshbach framework. Instead of applying the WKB
approximation for uncoupled fission barriers as often done in the
past, we solved the Schr\"{o}dinger equation for a one-dimensional
(1-D) potential model to calculate the penetration probabilities
(transmission coefficients) in the fission channel of compound nucleus
reactions. Because we took continuity of the fission path into
consideration, the expression to combine several penetrabilities for
different barriers, like $T = T_A T_B/(T_A+T_B)$, is no longer
involved in our model. Although the potential shape was parameterized
by smoothly concatenated parabolas for a sake of convenience, the
model can be applied to any arbitrary shape, as we obtain the wave
function by the numerical integration technique.

We showed that a resonance-like structure manifests in the calculated
transmission coefficients for the double-humped fission barrier that
includes a potential well between them, which is understood to be a
quantum mechanical effect in the fission channel. The resonance-like
structure becomes more remarkable when these barriers have a similar
height, where the penetration and reflection waves are in phase. The
structure becomes less sharp when an imaginary part is introduced in
the potential well. The complex potential also absorbs the flux of
fission channel, resulting in lower transmission coefficients.

The 1-D potential model was incorporated into the statistical
Hauser-Feshbach model to calculate neutron-induced reactions on
$^{235,238}$U and $^{239}$Pu. In this case we didn't include the
imaginary part in the potential. In order to calculate the potential
penetration for the excited states, we introduced a simple trajectory
compression model to account for change in the nuclear structure due
to the nuclear deformation. By aggregating the fission transmission
coefficients for all the possible fission paths that are energetically
allowed, calculated fission cross sections for $^{235,238}$U and
$^{239}$Pu were compared with the evaluated data that represent the
experimental cross sections.  We showed that reasonable reproduction
of the data can be obtained by a limited number of model
parameters. Although the detailed structure seen in the experimental
fission cross section is hardly reproduced by the 1-D model due to a
crude approximation for the potential adopted, further improvement
could be made by more careful studies on the potential shape, together
with more realistic trajectory compression models.

\section*{Acknowledgments}

TK thanks B. Morillon and P. Romain of CEA for valuable discussions on
this subject.  TK and PT were supported by the Advanced Simulation and
Computing (ASC) Program, National Nuclear Security Administration,
U.S. Department of Energy.  This work was carried out under the
auspices of the National Nuclear Security Administration of the
U.S. Department of Energy at Los Alamos National Laboratory under
Contract No. 89233218CNA000001.

%

\end{document}